\documentclass[reprint,twocolumn,showpacs,amsmath,amssymb,prl,ps]{revtex4-1}
\usepackage{graphicx,color}
\usepackage{amsthm}
\usepackage{array}
\usepackage{xy}
\usepackage{braket}
\usepackage{subfigure}
\usepackage{textcomp}
\usepackage{hyperref}
\usepackage[bottom]{footmisc}
\hypersetup{colorlinks=true, linkcolor=blue, citecolor=blue, filecolor=blue, urlcolor=blue, pdftitle=, pdfauthor=, pdfsubject=, pdfkeywords=,pdfstartview=FitH,}

\newcommand{\cdag}{c^{\dagger}}
\newcommand{\cnod}{c^{\phantom{\dagger}}}
\newcommand{\llangle}[0]{\langle\langle}
\newcommand{\rrangle}[0]{\rangle\rangle}
\newcommand{\lllangle}[0]{\langle\langle\langle}
\newcommand{\rrrangle}[0]{\rangle\rangle\rangle}

\allowdisplaybreaks 

\begin{document}

\title{Combined topological and Landau order from strong
  correlations in Chern bands}

 \author{Stefanos Kourtis}
 \author{Maria Daghofer}
 \affiliation{
  Institute for Theoretical Solid State Physics, IFW Dresden, 01171 Dresden, Germany 
 }

\date{\today}

\begin{abstract}
We present a class of states with both topological and
conventional Landau order that arise out of strongly interacting
spinless fermions in fractionally filled  and topologically non-trivial bands with Chern
number $C=\pm 1$. These quantum states show the features of
fractional Chern insulators, such as fractional Hall conductivity and
interchange of ground-state levels upon insertion of a magnetic flux. In
addition, they exhibit charge order and a related additional trivial
ground-state degeneracy. Band mixing and geometric frustration of the charge 
pattern place these lattice states markedly beyond a single-band description.
\end{abstract}

\maketitle

The fractional quantum Hall (FQH) effect is a paradigmatic example for a
correlation-driven state with topological order~\cite{Wen1990},
and the proposal~\cite{Neupert2011,tang10,sun10} that the concept
might be extended from Landau levels arising due to a magnetic field
to topologically nontrivial bands on a lattice has thus raised
considerable interest. 
The role of the magnetic
  field is then played by the band's Berry curvature and Coulomb repulsions are
  expected to stabilize analogues to FQH states.
That fractionally filled and topologically
non-trivial bands with non-zero Chern number $C$ can indeed host
corresponding states, dubbed fractional Chern insulators (FCI), has
since been well established
numerically~\cite{Neupert2011,Sheng2011,Regnault2011,Venderbos2011,Venderbos2011a}
for various models. 
Analytical
considerations have established connections to FQH states by proposing
wave functions~\cite{PhysRevLett.107.126803,PhysRevB.86.085129} and
pseudopotentials~\cite{2012arXiv1207.5587L}, as well as by noting the
close mathematical relation between density operators in partially
filled Chern bands and those of partially filled Landau
levels~\cite{PhysRevB.85.075128,PhysRevB.85.241308,PhysRevB.86.195146,chamon2012a}. As
paths to realizing topologically nontrivial and nearly flat
bands cold quantum gases~\cite{2012arXiv1212.3552C,2012arXiv1212.4839Y}, 
oxide interfaces~\cite{Xiao:2011} and
layered oxides with an orbital degree of
freedom~\cite{Venderbos2011,Venderbos2011a}, strained
graphene~\cite{PhysRevLett.108.266801}, and organometallic
systems~\cite{PhysRevLett.110.106804} have been proposed. A
review of this rapidly evolving field can be found in Ref.~\onlinecite{Parameswaran2013}. 

Apart from the intrinsic interest in such an effect, one motivation
for the search of FCIs is their energy and thus temperature
scale: as it is given by the scale of the interaction, it is expected
to  be  considerably higher than the sub-Kelvin range of the FQH
effect, especially for oxide-based
proposals~\cite{Xiao:2011,Venderbos2011,Venderbos2011a}. As one would
moreover not need strong magnetic fields, both realization of such
states and their potential application to qubits~\cite{Nayak2008}
then appear more feasible. It has been established that FCI states can
persist the influence of several aspects that make bands on lattices
different from perfectly flat Landau levels with uniform Berry curvature, e.g., finite
dispersion~\cite{Venderbos2011a,PhysRevB.86.205125}, a  moderate staggered chemical
potential~\cite{Neupert2011}, disorder~\cite{Yang2012,kourtis2012a},
or competition with a charge-density wave
(CDW)~\cite{kourtis2012a}.

Since the FQH effect can
be  discussed in tight-binding models instead of 
Landau levels~\cite{0953-8984-3-23-012,PhysRevLett.94.086803},
particularly intriguing features of
FCI states are those that go beyond their FQH counterpart. 
FCIs with higher Chern numbers were discussed~\cite{PhysRevB.86.205125,PhysRevB.86.201101,Trescher:2012tz}, which may
be non-Abelian and thus suitable for quantum computation. It has also been noted
that FCIs do not share the particle-hole symmetry of partially filled
Landau levels~\cite{lauchli2012,kourtis2012a}. 
All these extensions can,
however, be understood by focusing exclusively on the fractionally
filled Chern band. FCI states considered so far are weakly interacting,
in the sense that the interactions stabilizing them are too weak to mix in the other band(s) with
different Chern numbers. Those can be even projected out of
the Hamiltonian, which keeps the band topology intact but obscures the
impact of other aspects like lattice geometry, again
reflecting the similarity to  Landau levels with their weak lattice
potential.  

In this Letter, we go beyond the limit of `weak' interactions, into a
regime where Chern bands with $C=+1$ and $C=-1$ mix. We find
states that show the features of both a CDW (revealed by the 
charge structure factor) and of an FCI (fractional Hall conductivity
and spectral flow).  
The states are related to the `pinball liquid' of the triangular
lattice~\cite{Hotta2006} that combines charge-ordered with metallic~\cite{Nishimoto:2008fv,CanoCortes:2011fx}
or superconducting~\cite{Morohoshi:2008cm} electrons. In the latter
case, the system combines charge order with`off-diagonal
long-range order'. 
Simultaneous existence of two such different  
order parameters has been extensively investigated, especially in its bosonic
counterpart~\cite{PhysRevLett.95.237204,PhysRevLett.95.127205,PhysRevLett.95.127207,PhysRevLett.95.127206},
the supersolid~\cite{Boninsegni:2012hw}. 
%
The present case, however, differs fundamentally from supersolids or
superconducting pinball liquids, as the second type of order in
addition to charge order is \emph{topological}, i.e., non-local and
without an order parameter. We
thus arrive at an exotic state of matter that is characterized by both
Landau-type and topological order and is understood in terms of both band
topology and lattice geometry. 

\begin{figure}[t]
\centering
\includegraphics[width=\columnwidth]{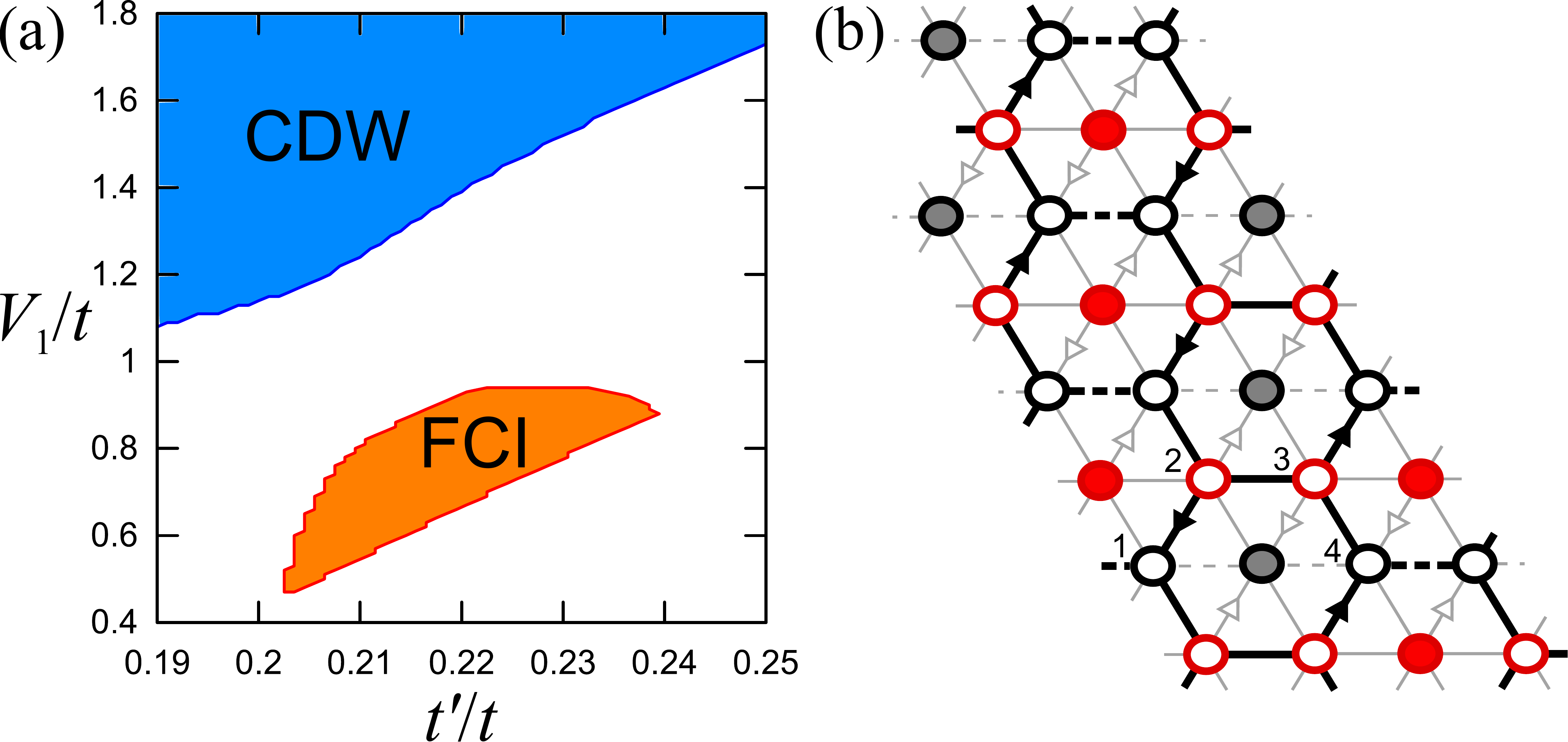}
\caption{(color online) 
CDW and FCI on the triangular lattice at
  $\nu=2/3$, resp. $\bar{n}=1/3$. (a) Phase diagram of model
  Eqs.~\eqref{eq:model} for a $3\times3$ unit-cell ($6\times 3$ lattice-site) system
  with 6 fermions and $V_2=V_3=0$. 
(b) The $4\times4-1=15$ unit-cell ($30$ lattice-site)
cluster used in this Letter, the unit cell (black/red circles),
the charge-order pattern (filled circles) arising in the CDW and the
hoppings. 
Solid, dashed and arrowed
black lines represent complex hoppings with phases of 0, $\pi$ and
$\pm\pi/2$ respectively. Grey lines denote hoppings deactivated by the
CDW. Third
nearest-neighbor hoppings are not shown.} 
\label{fig:phdiag23}
\end{figure}

This novel class of states can be intuitively understood as comprising of particles
that play two roles simultaneously. Most of them 
form the CDW occupying $1/3$ of the triangular lattice sites, see Fig.~\ref{fig:phdiag23}(b). 
As has been discussed in the context of the pinball
liquid~\cite{Hotta2006,Nishimoto:2008fv,CanoCortes:2011fx,Merino:2013wd},
additional particles can then move on the remaining sites, up 
to a total density of $\bar{n}=2/3$ (two fermions per three lattice sites). One can view the remaining sites as an
effective honeycomb-lattice model, which has here  a 4-site unit cell [see
Fig.~\ref{fig:phdiag23}(b)] and is described by a topologically
nontrivial hoppings. Residual as well as longer-range Coulomb
interactions will here be shown to induce an FCI out of the pinball liquid's metal, the {\it topological pinball liquid} (TPL).

We discuss here spinless fermions on a triangular lattice. 
The topologically nontrivial kinetic energy can be expressed in momentum space as 
\begin{subequations}
\begin{align}\label{eq:ekin}
 {\cal H}^0 &= \sum_{{\bf k},\mu,\nu} \cdag_{{\bf k},\mu} H^0_{\mu,\nu}({\bf k}) \cnod_{{\bf k},\nu}
\end{align}
where the indices $\mu,\nu$ refer to a two-site unit
cell and $\cdag_{{\bf k},\mu}$ ($\cnod_{{\bf
    k},\mu}$) are fermion creation (annihilation) operators. The
momentum dependence is contained in the $2\times
2$ matrix $H^0({\bf k})$ expressed in terms of the vector of isospin Pauli matrices $\boldsymbol\tau$ and the unity
matrix $\hat I$ as 
\begin{align}
H^0({\bf k}) &= {\bf g(k)}\cdot{\boldsymbol\tau} + g_0({\bf k}) \hat I ,\ \textrm{with}\\
 g_i({\bf k}) &=  2t \cos({\bf k}\cdot{\bf a}_i), \ i=1,2,3 \ \textrm{and} \\
 g_0({\bf k}) &= 2t' \sum_{i=1}^3 \cos(2{\bf k}\cdot{\bf a}_i) . \label{eq:mat_k}
\end{align}
The unit cell and the phases of the hoppings can be seen in
Fig.~1(b). 
${\bf a}_1 = (1/2, -\sqrt{3}/2)^T$, ${\bf a}_2 = (1/2, \sqrt{3}/2)^T$
and ${\bf a}_3 = -( {\bf a}_1+{\bf a}_2 )$ are the triangular-lattice
unit vectors. The nearest-neighbor (NN) and third nearest-neighbor
hopping matrix elements are $t$ and $t'$, the latter is tuned to
change the dispersion of the Chern band~\cite{kourtis2012a}, with
flattest bands arise for $t'/t \approx 0.2$. We keep $t'/t = 0.2$ 
unless mentioned otherwise, but we have verified that the main results remain
valid for other values. 
The interaction
\begin{align}
 {\cal H}^1 = V_1 \sum_{\langle i,j \rangle} \hat n_i \hat n_j 
+ V_2 \hspace{-0.25em}\sum_{\llangle i,j \rrangle} \hat n_i \hat n_j
+ V_3\hspace{-0.35em} \sum_{\lllangle i,j \rrrangle} \hat n_i \hat n_j, \label{eq:ecoul} 
\end{align}\label{eq:model}
\end{subequations}
is a repulsion of strength $V_1$ between NN sites, denoted by $\langle
i,j \rangle$ and longer-range repulsion $V_2$ and $V_3$ between
second and third neighbors, respectively. We present here results for $V_2=V_3$,
but have verified that unequal values do not qualitatively change the
results unless very strong $V_2$ destabilizes the $V_1$-driven
CDW. Operator $\hat n_i$ measures particle density at site $i$. 
While the model was originally introduced to describe topologically
nontrivial phases in Kondo-lattice~\cite{Martin:2008dx} and
$t_{2g}$~\cite{Venderbos2011a,kourtis2012a} systems -- hence the signs and phases
in Eq.~\eqref{eq:model} -- we use it here to study
the more generic question of FCI states in the limit of strong interactions. 
We treat the model Eqs.~\eqref{eq:ekin} and \eqref{eq:ecoul} with
Lanczos exact diagonalization on small clusters with periodic
boundaries directly in real space, without projection onto the single-particle
bands - both bands with Chern numbers $C=\pm 1$ are kept. We use
clusters with $3\times 6$ and 30 sites, which corresponds to 9
resp. 15 unit cells, complemetned by larger clusters in the limit of large $V_1$. 

The model of Eqs.~\eqref{eq:model} has been shown to yield FCI states
at several filling fractions~\cite{Venderbos2011a,kourtis2012a}. 
As the proposed TPL states we will be discussing below
have both charge order and topological order, let us 
first briefly review these two phases and their signatures.  
For a filling of $\bar{n} = 1/3$
and small to moderate $V_1$, the electrons occupy $\nu=2/3$ of 
the lower band and a corresponding FCI arises. CDW fluctuations
destroy it at larger $V_1$, see the phase diagram Fig.~\ref{fig:phdiag23}(a) and
Ref.~\onlinecite{kourtis2012a}.  A schematic picture of the CDW is
given in Fig.~\ref{fig:phdiag23}(b).

Appropriate groundstate properties are necessary to determine the precise
nature of the various phases. 
Charge order shows up in the 
charge-structure factor 
\begin{align}\label{eq:nk}
N({\bf k}) = \Bigl| \sum_{i} \textrm{e}^{i{\bf k}{\bf r}_i} (\hat{n}_i -\bar{n})
|n\rangle \Bigr|^2\;,
\end{align}
where $|n\rangle$ is a vector of the ground-state manifold and ${\bf
  r}_i$ denotes the location of site $i$.  
The CDW with $\bar{n}=1/3$ induces sharp peaks at momenta ${\bf
  k}=\pm {\bf K}$, where ${\bf K}=(2\pi/3,0)$, that grow with interaction strength $V_1$. In the FCI, their
weight should remain comparable to that of other
momenta~\cite{kourtis2012a}.

FCI states, on the other hand, are identified by a fractional Hall conductivity $\sigma_{\textrm{H}}$, which
is obtained by integrating the many-body Berry curvature in the
$\boldsymbol\varphi = (\varphi_{{\bf a}_2},\varphi_{{\bf a}_3})$-plane of magnetic fluxes through the handles of
the torus, 
where the fluxes are introduced as 
phase factors in the hoppings along ${\bf a}_2$ and ${\bf a}_3$,
respectively. 
It is evaluated with the Kubo formula~\cite{Niu1985,Xiao2010,kourtis2012a}: 
\begin{equation}
 \sigma_{\textrm{H}} = \frac{N_c}{\pi q} \sum_{n=1}^q \iint_0^{2\pi} d\varphi_{{\bf a}_2} d\varphi_{{\bf a}_3} \Im \sum_{n'\not=n} \frac{ 
   \langle n | \frac{\partial { H}}{\partial \varphi_{{\bf a}_3}} |
     n' \rangle 
   \langle n' | \frac{ \partial { H}}{ \partial
     \varphi_{{\bf a}_2}} | n\rangle}
{(\epsilon_n -
   \epsilon_{n'})^2},\label{eq:chern}
\end{equation}
where $N_c$ is the number of unit cells and $\ket{n'}$ are higher-energy
eigenstates with eigenenergies
$\epsilon_{n'}$. $\epsilon_{n}$ are the energies of the ground states.  All values of $\sigma_{\textrm{H}}$ are given in units of
$e^2/h$. As this approach does not involve projections onto
the lower (flat) band, it remains valid for arbitrary interactions
and band mixing. 
In the FCI regime, we find a very precisely quantized 
$\sigma_{\textrm{H}} = 2/3 $. 
For the $3\times 6$-sites  cluster, the 
ground states have the same momentum and the lowest-energy
state contributes $2$, while the 
other two states do not conduct, giving the expected average. 

In both a CDW with $\bar{n}=1/3$ and an FCI with $\nu=2/3$, we expect a gapped
ground-state manifold of three nearly degenerate states that are
separated from the remaining spectrum by a gap. If their momenta are different, the
FCI ground states are expected to exhibit spectral flow upon insertion of one
flux quantum through one of the handles of the torus~\cite{Tao1984,Thouless1989}.
The closing of the gap for any value of the fluxes determines the phase boundaries
in Fig.~\ref{fig:phdiag23}(a).

Using all eigenvalue and ground-state properties discussed, we trace
the approximate phase diagram  of the Hamiltonian given by
Eqs.~\eqref{eq:model} in Fig.~\ref{fig:phdiag23}(a), setting  $V_2=V_3=0$.
Comparison to earlier results~\cite{kourtis2012a} for a lattice that is not commensurate with the
CDW and hence suppresses it, reveals substantial finite-size effects, but
the presence and approximate location of both phases are consistent: The stability of the 
FCI depends both on $V_1$ and on the hopping $t'$; it is most stable for $t'/t\gtrsim
0.2$. At $\bar{n}=1/3$, the CDW can always be induced by increasing $V_1$ and
there is no coexistence of FCI and CDW phases.  

\begin{figure}[t]
\centering
\subfigure{\includegraphics[width=0.49\columnwidth, trim=0 0 0 0,clip]{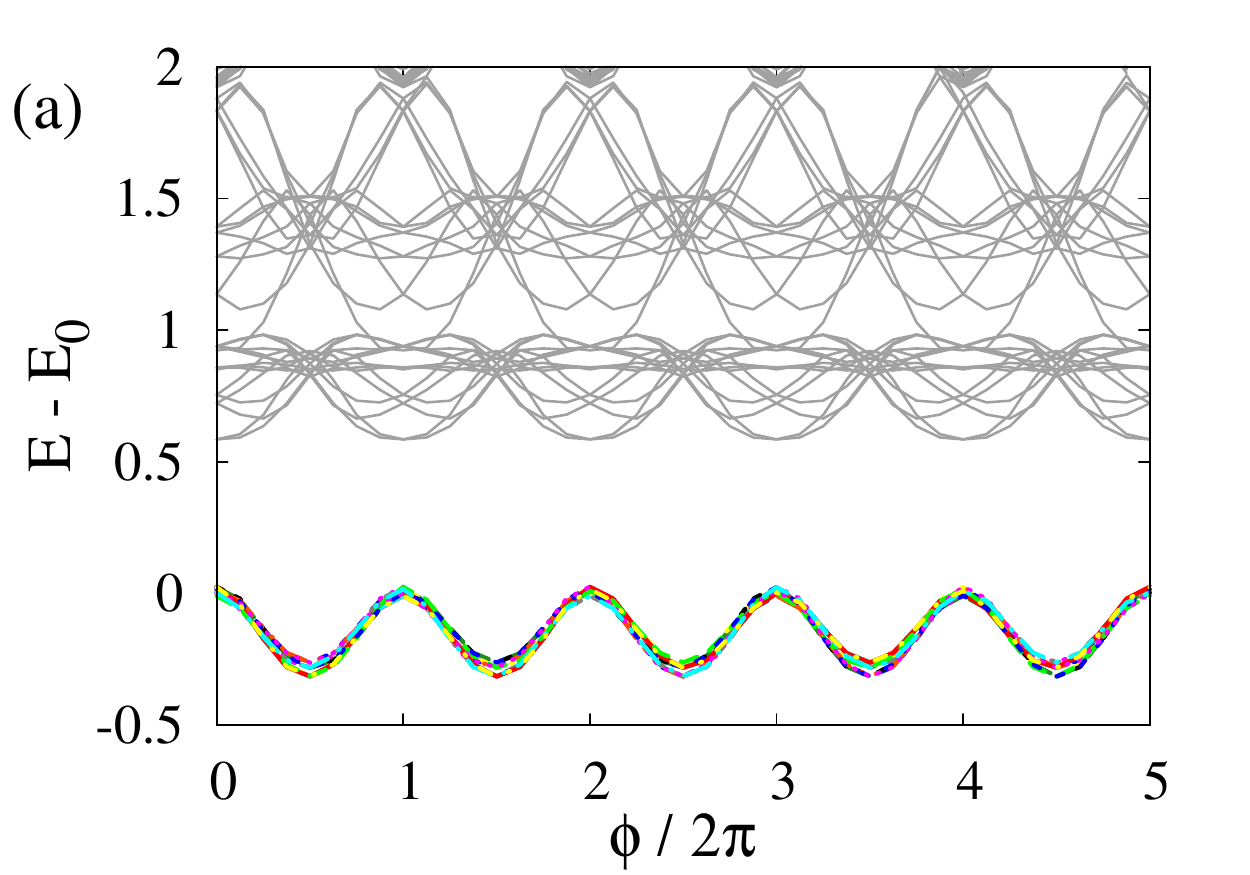}\label{fig:evflux_12}}
\subfigure{\includegraphics[width=0.49\columnwidth, trim=0 0 0 0, clip]{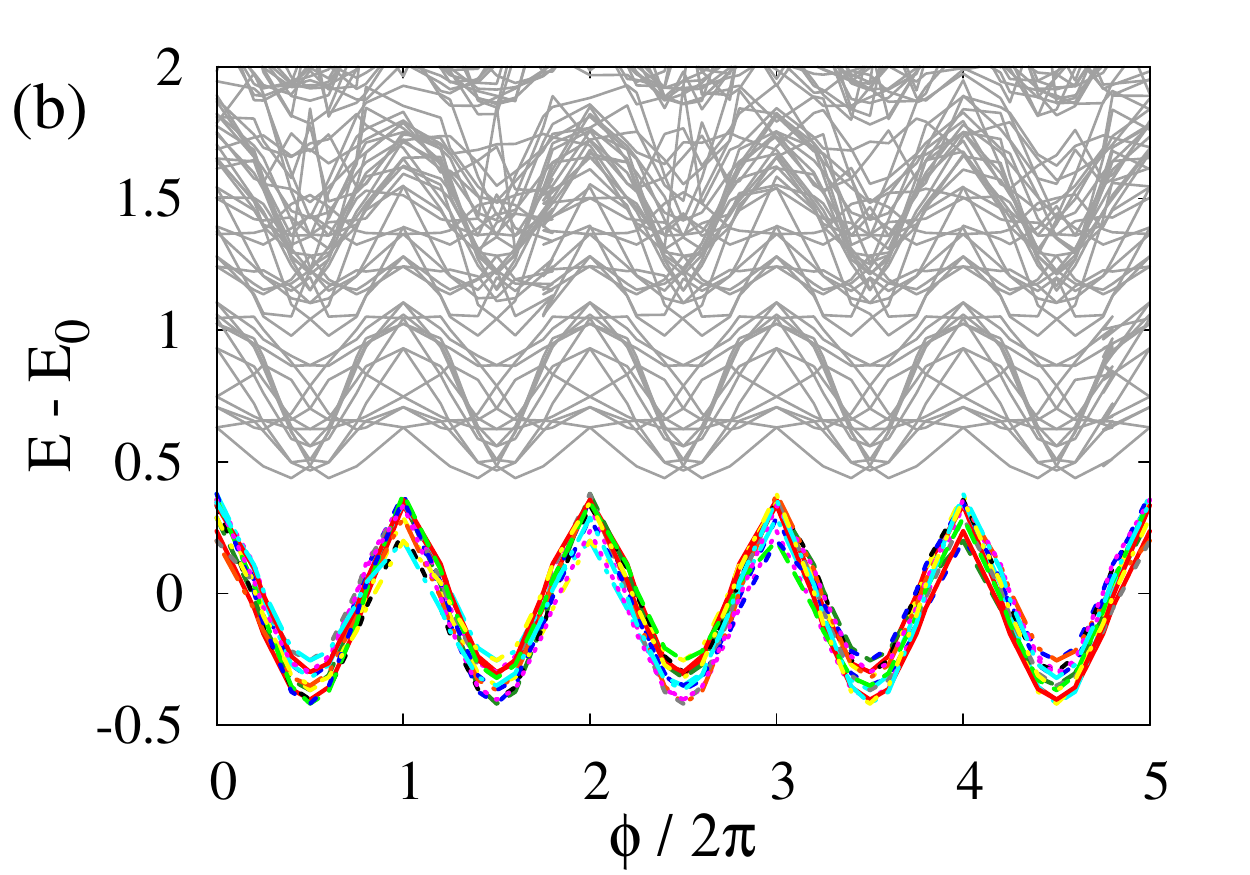}\label{fig:evflux_13}}\\
\begin{minipage}[b]{0.49\columnwidth}
\includegraphics[width=\textwidth, trim=0 45 0 0, clip]{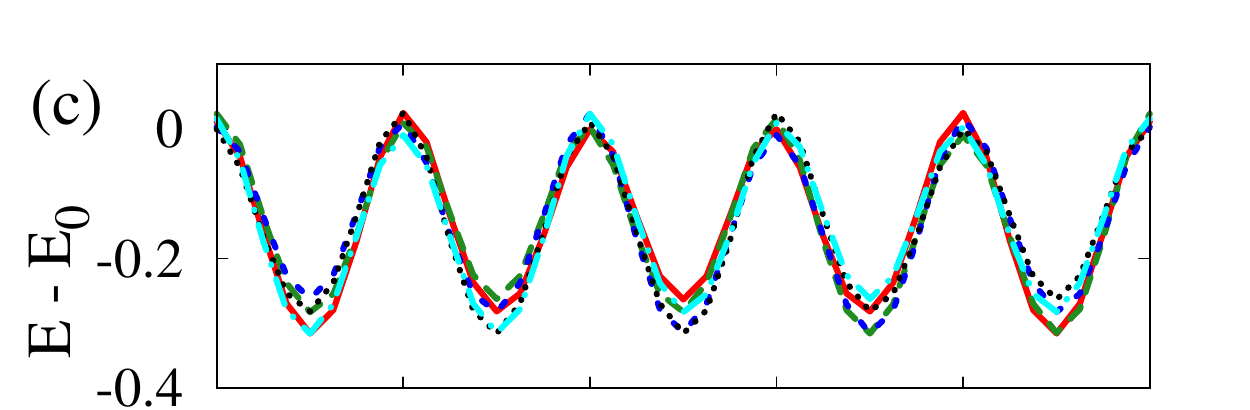}\\[-0.2em]
\includegraphics[width=\textwidth,trim=0 45 0 0,clip]{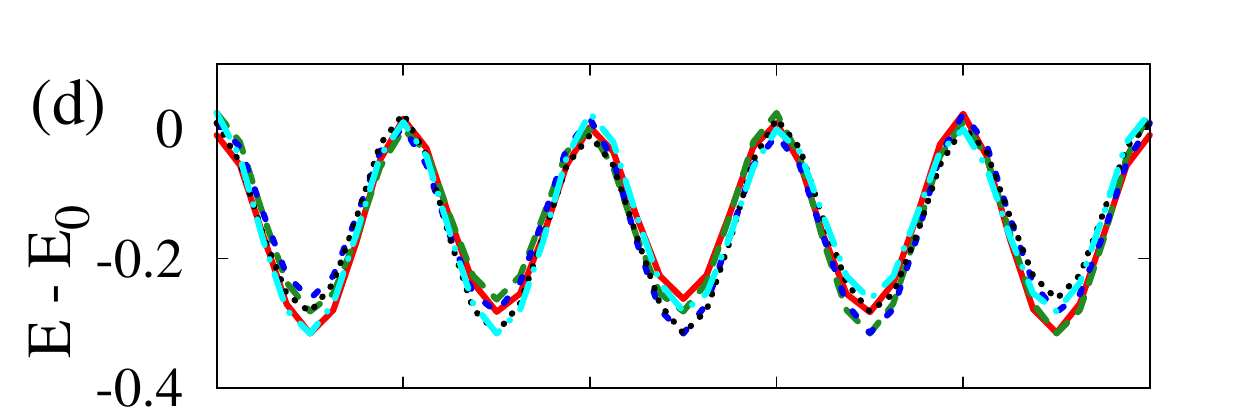}\\[-0.2em]
\includegraphics[width=\textwidth,trim=0 0 0 0,clip]{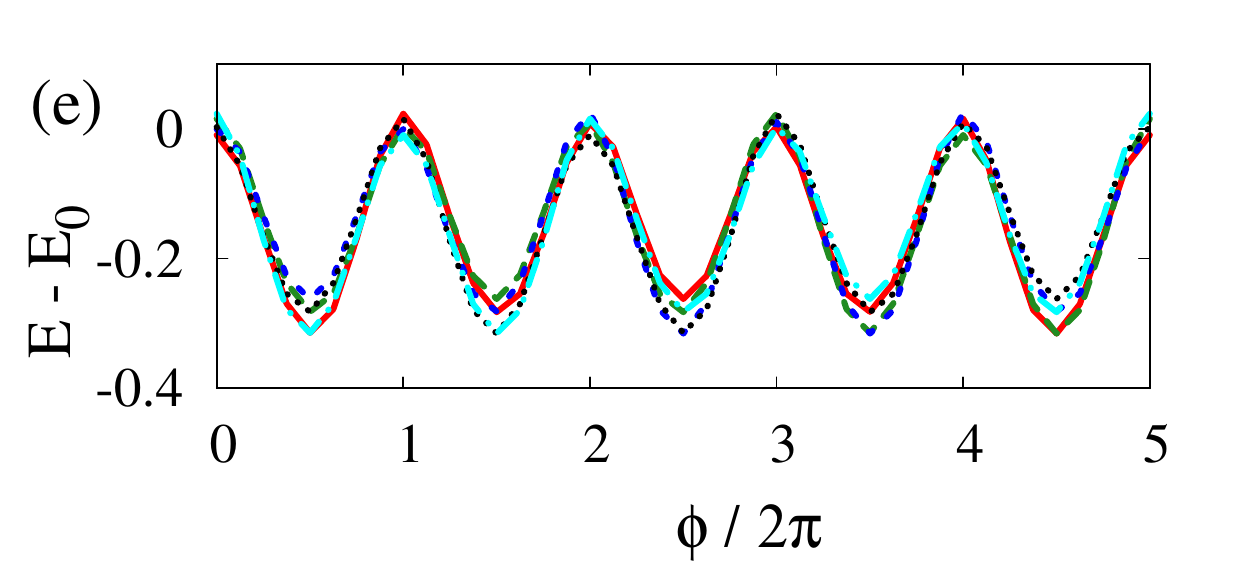}\\
\end{minipage}
\begin{minipage}[b]{0.49\columnwidth}
\includegraphics[width=\textwidth, trim=0 45 0 0, clip]{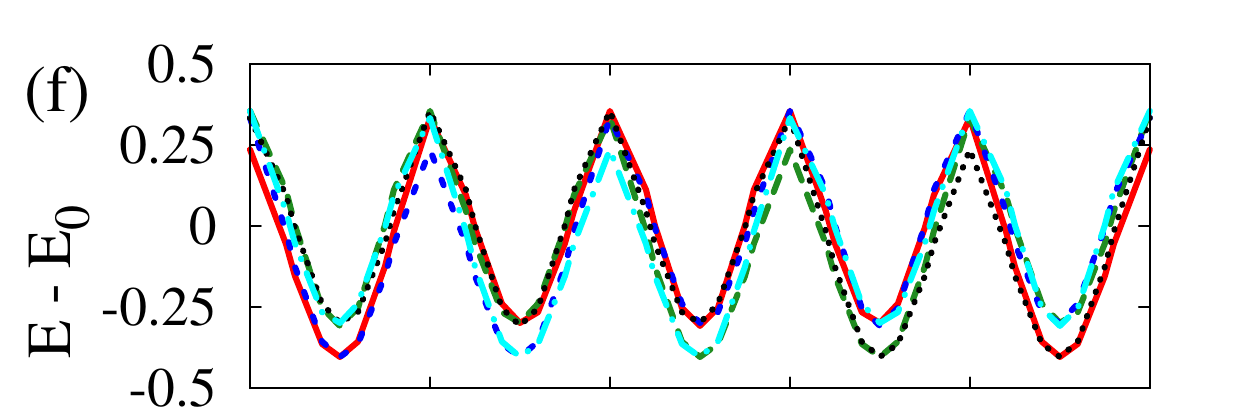}\\[-0.2em]
\includegraphics[width=\textwidth,trim=0 45 0 0,clip]{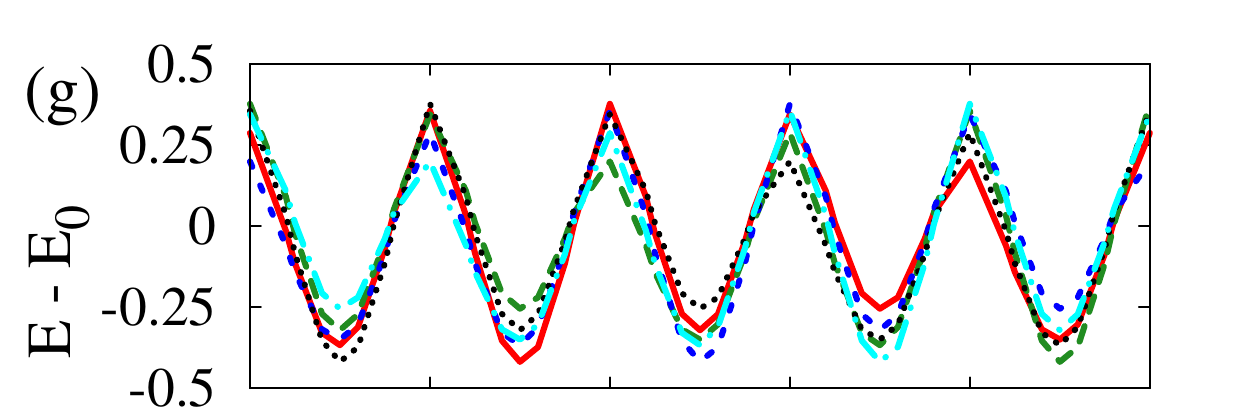}\\[-0.2em]
\includegraphics[width=\textwidth,trim=0 0 0 0,clip]{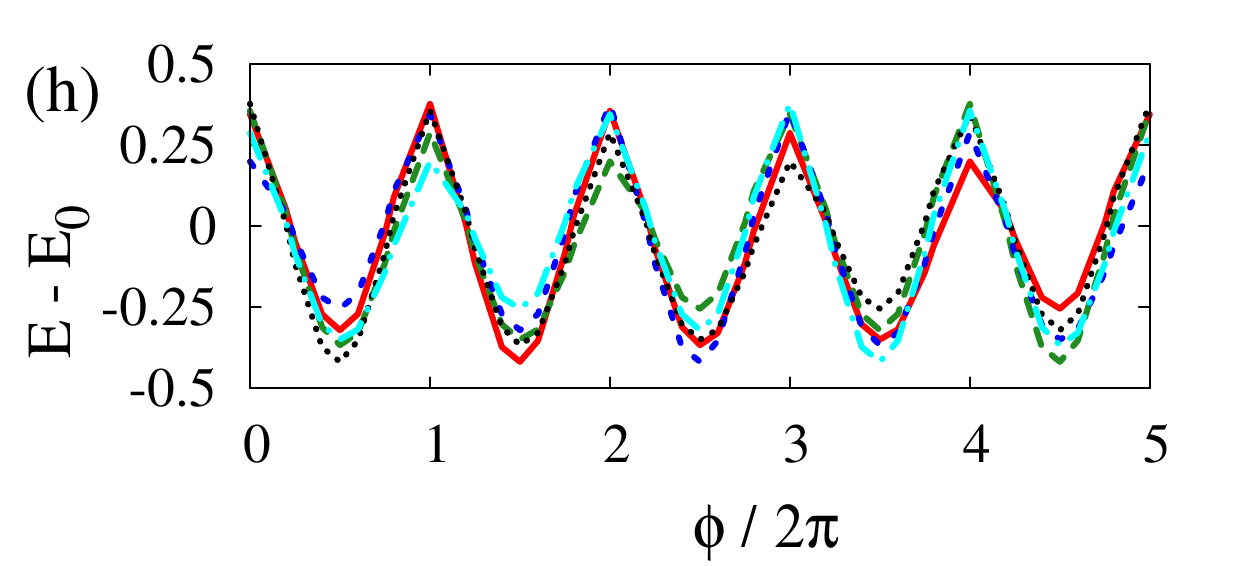}\\
\end{minipage}
\caption{(Color online) Spectral flow upon flux insertion for the
  model of Eqs.~\eqref{eq:ekin} and \eqref{eq:ecoul} for a $15$
  unit-cell system at (a) $\nu=12/15$ and (b) $\nu=13/15$. In (c-e)
  and (f-h),
  the 15 low-energy states of (a) resp. (b) are divided into three groups of
  five; each group shows spectral flow consistent with a denominator-5
  FCI. Fluxes are inserted as additional phases in the  
  hoppings, totaling to $\varphi$ for a loop around the cluster
  along direction ${\bf a}_3$. Parameters are $t'/t=0.2$, $V_1/t=10$ and
  $V_2=V_3=2t$.} 
\label{fig:evflux}
\end{figure}

Having recognized the main features of the FCI, the CDW, and their
competition at $\bar{n}=1/3$, we now turn to a phase having the defining
features of both. Figure~\ref{fig:evflux} gives the eigenvalue
spectrum for fillings $\bar{n}=12/30, 13/30$ and shows a 15-fold
degenerate ground state as well as spectral flow. 
For the first case,
the degeneracy expected for a straightforward FCI with $\nu =12/15=4/5$
would be five, which we did not observe for any interaction strength,
instead of 15.  In  the second case, the degeneracy is 
consistent with a $\nu=13/15$ FCI, however, levels return to their initial
configuration already after insertion of only five flux quanta, which
is unexpected. 
Moreover, additional interaction-generated dispersion tends to
destabilize FCIs  at such high fillings~\cite{chamon2012a,lauchli2012}. 
Closer inspection shows that the 15 low-energy states 
can be separated into three groups of five states each,
where each group shows the spectral flow expected for a
denominator-five state, see Figs.~\ref{fig:evflux}(c-e) and~\ref{fig:evflux}(f-h). 

The Hall conductivity establishes this similarity to $\nu =2/5$ ($\nu
= 3/5$) states rather than $4/5$ ($13/15$), as it is precisely quantized to
$\sigma_{\textrm{H}} = 0.4$ ($\sigma_{\textrm{H}} = 0.6$)
for each of the ground states. 
As in some FCI states, e.g. $\nu = 2/3$, the sum of contributions to
the 
Hall conductivity does not add up to the Chern number of the
non-interacting band. 
However, the present states are considerably more exotic as
$\sigma_{\textrm{H}}=m/n$ at $\nu=p/q$, with  $n\not=q$. 
The Hall conductivity is thus not
given by the usual heuristic $\sigma_{\textrm{H}}=\nu \times C$. 
Similar to earlier observations for the FQH effect in the presence of an external
potential~\cite{Kol1993}, this is a strong indication that the
`topological' degeneracy differs from the number of ground states and
is here in both cases $n=5$ rather than $q=15$.

\begin{figure}[t]
\centering
\subfigure{\includegraphics[width=0.99\columnwidth]{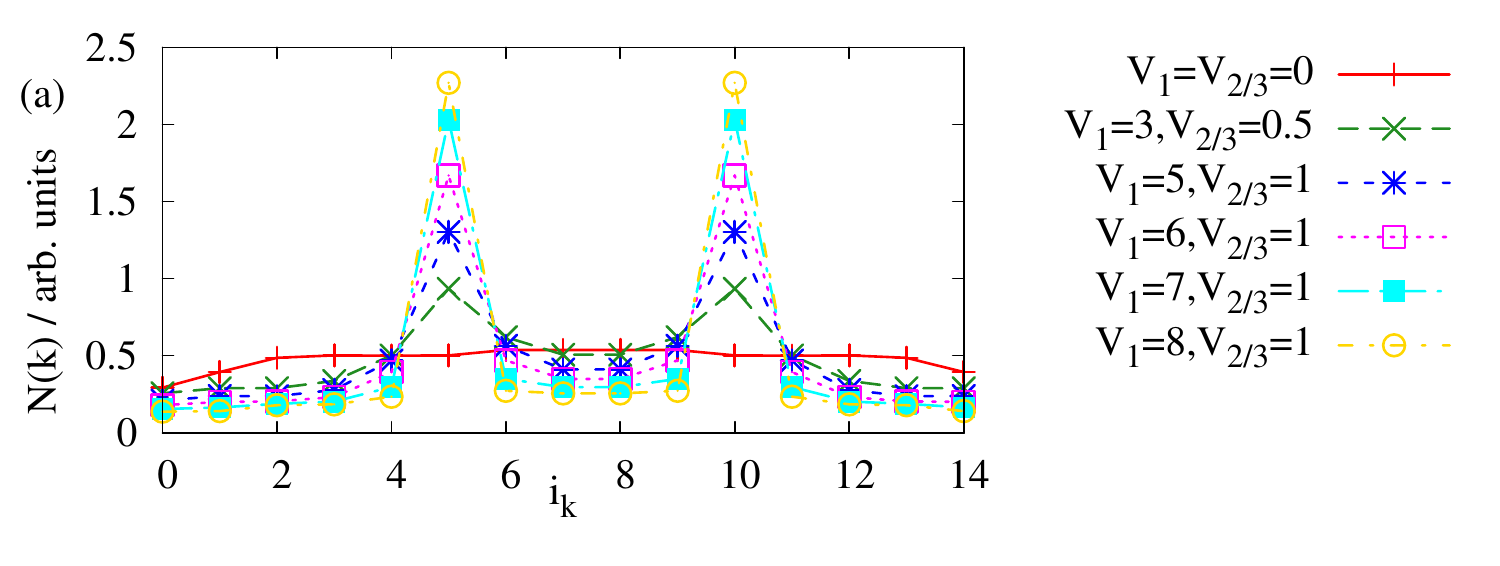}\label{fig:nk_12}}\\[-2.5em]
\subfigure{\includegraphics[width=0.425\columnwidth]{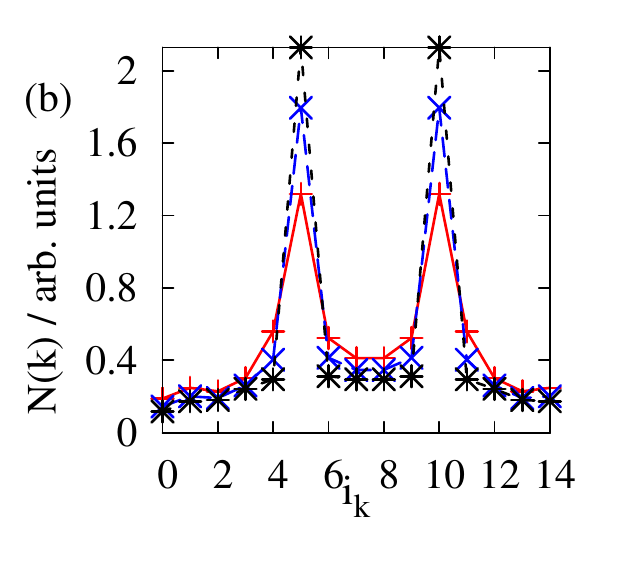}\label{fig:nk_13}}
\subfigure{\includegraphics[width=0.555\columnwidth]{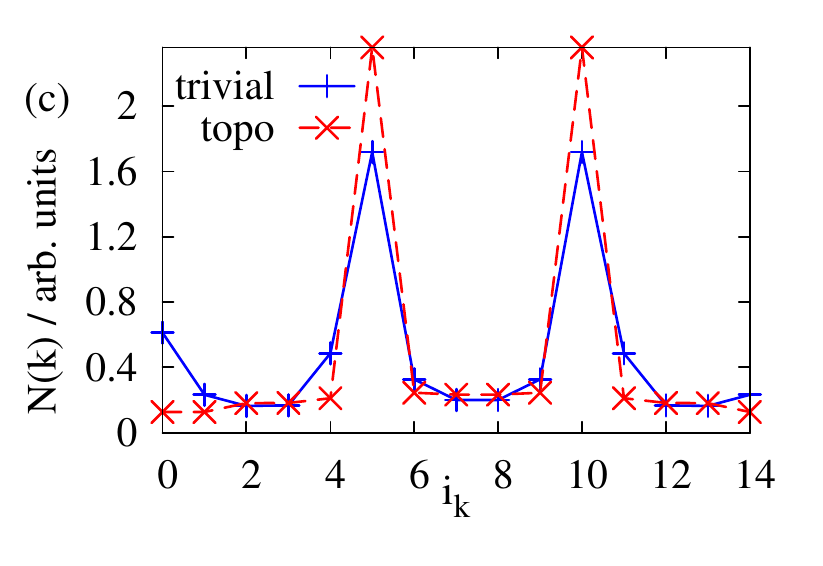}\label{fig:nk_rest}}
\caption{(Color online) Charge-structure factor $N({\bf k})$.
(a) for $\bar{n}  = 12/30$ and (b) for $\bar{n}  = 13/30$, $V_2=V_3=2t$, and
$V_1=8t$ ($+$), $V_1=10t$ ($\times$), and $V_1=12t$
($\ast$). (c) Comparison of topologically trivial
  and nontrivial kinetic energy  for $V_1/t=10$,
  $V_2=V_3=2t$. $t'=0.2t$ and in the trivial case, $g_3=m=2t$~\cite{Note2}
Of the 15 available momenta, numbers 5 and 10 correspond the ordering
  momenta $\pm {\bf K}$ of the CDW. 
\label{fig:nk}}
\end{figure}

The remaining `trivial' three-fold degeneracy stems from Landau-type
charge order, 
as revealed by the static
charge-structure factor Eq.~(\ref{eq:nk}) depicted in Figs.~\ref{fig:nk_12}
and~\ref{fig:nk_13}.  For $V_1>0$, it peaks at $\pm {\bf K}$; the peaks grow when
stronger $V_1$ enhances charge order. As discussed above, this CDW has three
quasi-degenerate ground states.
We thus conclude that we have five FCI states (corresponding to $\nu=2/5$
or $\nu=3/5$) for each of the three ground states of the CDW, totaling to the 15
ground states. Five FCI states per CDW state exhibit
spectral flow, as seen in Fig.~\ref{fig:evflux}, 
and return to the original point after insertion of 5 fluxes. 
The CDW itself does not rely
on band topology: 
Going to a topologically trivial mass term $g_3=2t=\textrm{const.}$~\cite{Note2}
somewhat favors a competing sublattice ordering, but charge
order nevertheless remains strong,  see
Fig.~\ref{fig:nk_rest}. However, the state is then topologically trivial
with $\sigma_{\textrm{H}}=0$.  

\begin{figure}[t]
\centering
\includegraphics[width=0.7\columnwidth]{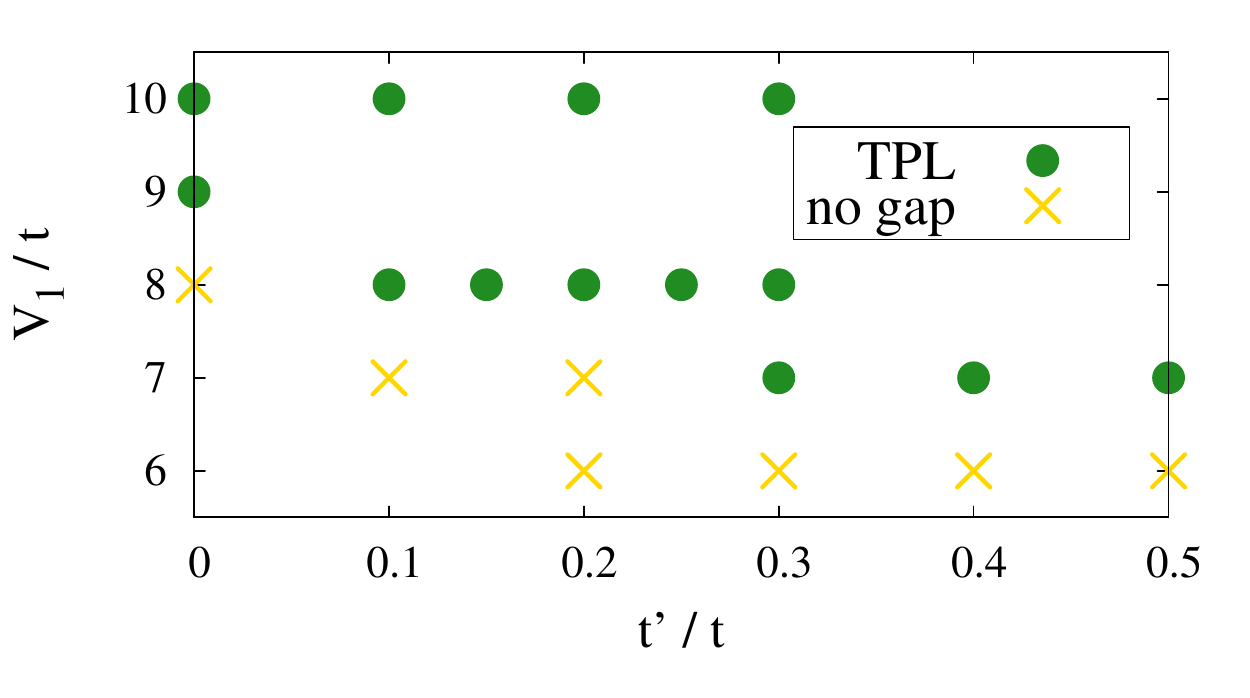}\\[-1.5em]
\caption{(Color online) Phase diagram for $\bar{n}  = 12/30$ 
depending on $V_1$ and $t'$ for $V_2=V_3=2t$.
Filled circles denote  TPL states, $\times$ denotes 
states that do not show a gap between the lowest 15 states and
the rest of the spectrum for all fluxes $\boldsymbol\varphi =
(\varphi_{{\bf a}_2},\varphi_{{\bf a}_3})$~\cite{suppl}.
\label{fig:phases}}
\end{figure}

Comparison of the phase diagram 
Fig.~\ref{fig:phases}  to Fig.~\ref{fig:phdiag23}(a)
shows that
the TPL needs stronger interactions than the simple CDW at $\bar{n}=1/3$,
consistent with observations in supersolids~\cite{PhysRevLett.95.127205}.
In contrast to the pure FCI, the TPL is not induced more easily for
the nearly flat bands at $t'\approx 0.2t$ than for more dispersive
bands. 
Moreover, 18 electrons (i.e. 12 holes) also show a
   TPL with charge order, 15 ground states, and $\sigma_{\textrm{H}}=-0.4$, even
  though the ``lower'' band for holes is quite dispersive~\cite{suppl}.
This may be connected to an intrinsically
reduced dispersion in the pinball state~\cite{Merino:2013wd}, or may be due to the rather
strong interactions needed to stabilize a CDW, which can then overcome
substantial dispersion~\cite{kourtis2012a}.  Interactions,
together with band topology and a partial frustration of the CDW, dominate
here over the details of the lower Chern band that were important
at weaker interactions. We also note that the cartoon picture of Fig.~\ref{fig:phdiag23}(b)
does not fully capture the correlated quantum
character of the TPL: for perfect charge order
and $t'=0$, the lowest subband of the effective system is not a Chern
band, yet, charge fluctuations at finite $V_1$ allow a TPL, see
Fig.~\ref{fig:phases}. 

Nevertheless, we can exploit the cartoon - valid in the limit of
strong $V_1$ - to
address larger systems, see the Supplemental Material. We restrict
the Hilbert space to low-energy states,  i.e., we remove states with too
many pairs of particles occupying NN bonds and paying $V_1$. For $V_1
\gg t$, the CDW becomes perfect and remaining particles move on the effective honeycomb lattice of
Fig.~\ref{fig:phdiag23}(b). Up to 6 particles moving on a 60-site 
honeycomb model then become accessible. (Corresponding to 90 triangular-lattice sites, of
which 30 are occupied by the CDW.) Indeed, the signatures of the
$\nu=2/5$ FCI component  are found, as expected for a TPL with  $\nu=12/15$.

In conclusion, we have presented numerical evidence for a class of
composite states of spinless fermions exhibiting both Landau and
topological order. The eigenvalue spectra of our interacting
spinless-fermion model on the triangular lattice at filling fractions
$\nu=12/15=4/5$ and $13/15$ hint at ground states that are neither FCI nor
CDW, but have features of both. The Landau order, commensurate charge 
modulation in the ground states, reveals itself in the interaction strength-dependent
peaks in the static charge-structure factor. The topological order is
established via the Hall
conductivity, which is precisely quantized, but with a value $\sigma_{\textrm{H}}\not=\nu$. Instead, we find a
quantization consistent with viewing the ground states as composites
of a CDW state and a FCI state formed by additional particles 
in the part of the lattice that remains unoccupied by the CDW, in some
sense similar to the superfluid coexisting with a CDW in 
a supersolid. 

These states with coexisting Landau and topological order mix both
bands of the model, with Chern numbers $C =\pm 1$, 
and are made possible by the geometric frustration of the triangular
lattice. 
The TPL is thus a state that arises out of lattice features that go beyond
the single-band picture usually sufficient to describe FCIs
and definitely goes beyond a Landau-level description. In contrast to
FCIs arising in a magnetically ordered system~\cite{Neupert:2011ee,Venderbos2011a}, where Landau order is
found in a different degree of freedom, both types of order are
here in the charge sector. In analogy to
the supersolid/pinball liquid connection on the triangular lattice, similar
states might be found on other lattices supporting
supersolids or pinball liquids, possibly also in multi-orbital
settings~\cite{Trousselet:2012kc}.

\begin{acknowledgments}
This work was supported by the Emmy-Noether program of the Deutsche
Forschungsgemeinschaft (DFG). We thank A.~Ralko, J.~Venderbos,
I. Vincon, T.~Neupert and P.~Kotetes for helpful discussions.
\end{acknowledgments}



%

\clearpage
\appendix

\chapter{\bf \large Supplemental Material for: Combined topological and Landau order from strong
  correlations in Chern bands}

\begin{figure}[hb]
\centering
\includegraphics[width=0.99\columnwidth]{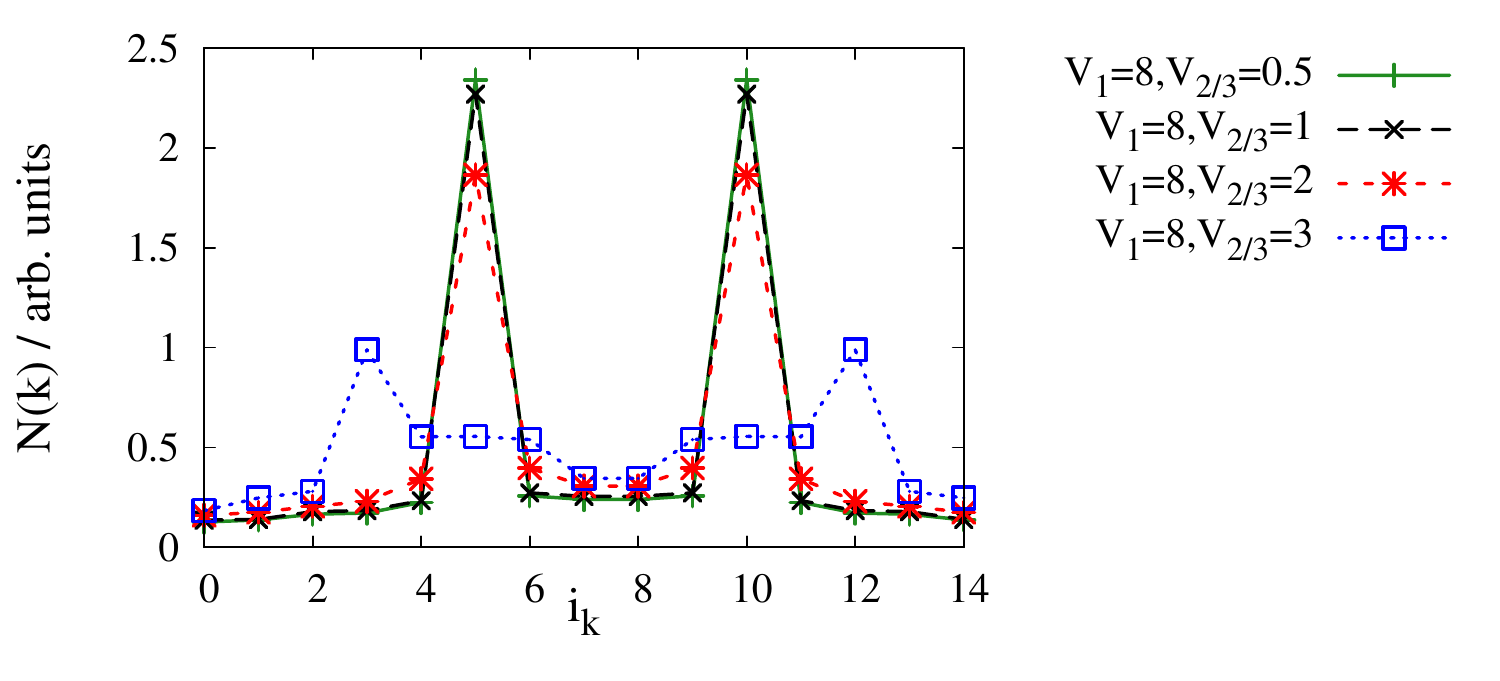}
\caption{(Color online) 
Static charge-structure factor $N({\bf k})$ illustrating how 
longer-range coulomb repulsion $V_2=V_3$ destabilizes the CDW pattern
shown in Fig. 1(b) of the main text. Momenta with numbers 5 and 10 correspond to $\pm {\bf K}$, the ordering
  momentum of the CDW, for $V_2=V_3 =3 t$, $N({\bf k})$ is no longer
  peaked here.  $V_1=8t$ and $t'=0.2t$ in all cases, filling is 12
  electrons on 30 sites/15 unit cells.
\label{fig:nk_V2}}
\end{figure}

\begin{figure}[hb]
\centering
\subfigure{\includegraphics[width=0.5\columnwidth]{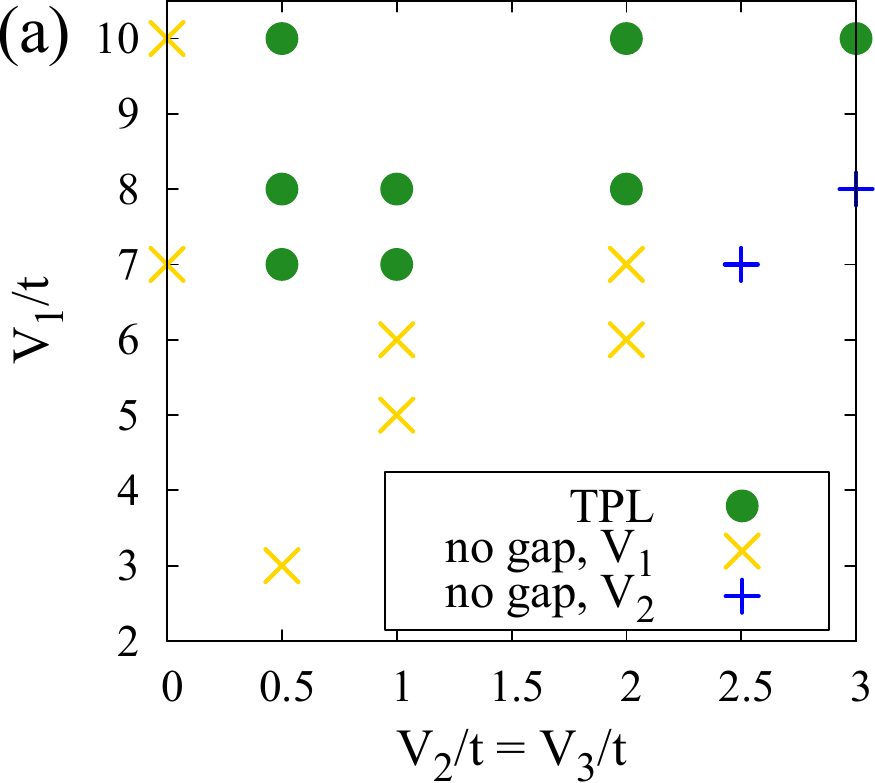}}
\quad \subfigure{\includegraphics[width=0.44\columnwidth]{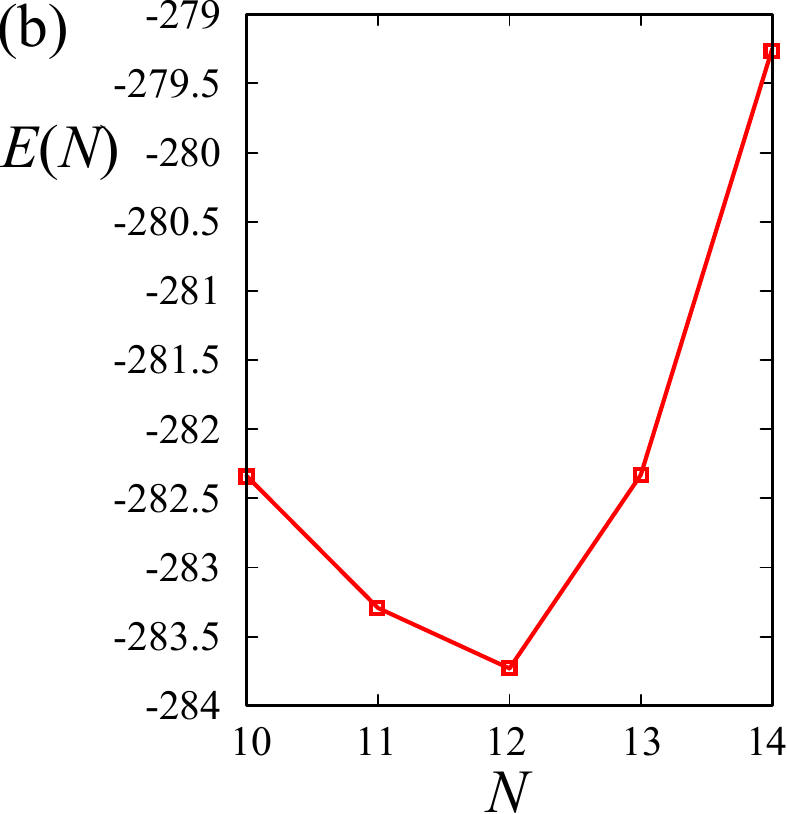}\label{fig:ps}}
\caption{(Color online) (a) Phase diagram for $\bar{n}  = 12/30$ depending
  on $V_1$ and $V_2=V_3$ for $t'=0.2t$. Filled circles denote  TPL
  states. $\times$ denote non-TPL states, which do not show a gap between the lowest 15 states and
the rest of the spectrum for all fluxes $\boldsymbol\varphi =
(\varphi_{{\bf a}_2},\varphi_{{\bf a}_3})$, but where $N({\bf k})$ is
still peaked at $\pm {\bf K}$.  $+$ denote non-TPL states where these peaks
in $N({\bf k})$ have been lost, i.e., where $V_2$
destabilizes the charge pattern, see Fig.~\ref{fig:nk_V2}.
(b) Total energy as a function of particle number on
the 30-site cluster for $V_1/t=10, V_2/t=V_3/t=2$.}
\label{fig:phases_V23}
\end{figure}

Figure~\ref{fig:nk_V2} shows that $V_2=V_3 \gtrsim 3t$ destroys the
CDW depicted in Fig, 1(b) of the main text for $V_1=8t$. Generally,
the critical value is $V_2\approx V_1/3$. As can be seen in the phase
diagram for $t'=0.2t$ in Fig.~\ref{fig:phases_V23}(a), finite $V_2$ and
$V_3$ is helpful, and may even be necessary, in inducing the TPL. When
longer-range interactions are strong enough to weaken the CDW,
however, first the TPL disappears and $N({\bf k})$ finally is no longer
peaked at $\pm {\bf K}$.

Phase separation into a charge
ordered and an FCI domain within the lattice would be hard to
reconcile with the observables discussed in the main text. Moreover,
we checked that the total energy depending on filling is convex for fillings
between 10 and 14 electrons on the 30-site cluster, see
Fig.~\ref{fig:ps}, also arguing for a
thermodynamically stable phase. 

For a filling of 18 electrons on 30 sites, we likewise found (i) charge order,
(ii) a 15-fold near ground-state degeneracy, and (ii)
$\sigma_{\textrm{H}}=-0.4$ (with $V_1/t=20$,
$V_2/4=V_3/4=4$, $t'/t=0.2$). This agrees with expectations based on a
particle-hole transformed situation of the 12-electron case discussed
in the main text: 10 holes form the CDW while the remaining 2 move
between them. This supports our picture of the topological pinball
liquid, as the ``normal'' pinball liquid is stable for $1/3< \bar{n} <
2/3$ and extends the stability range of the TPL.

While a finite-size scaling involving
several cluster sizes at the same filling is not possible in our case,
we were able to address somewhat larger clusters for the limit of strong
nearest-neighbor Coulomb repulsion. In this limit, the Hilbert space
can be reduced by discarding high-energy states with too many
electrons occupying NN sites, i.e., only states with an `almost perfect' CDW
have to be kept. We can then treat a cluster with $6\times 6$ sites
(18 unit cells) and a filling of 14 electrons. We allowed 0, 1, or 2
faults in the CDW, by finding consistent results, we can conclude that
the approximated Hilbert space still captures all relevant states. In
agreement with the expectations for a topological pinball liquid, we
find a nine-fold degenerate ground state, 3-fold for the CDW and
3-fold for the FCI. The latter can be understood in terms of two
electrons that fill one third of the lowest subband resulting from the
effective lattice. Accordingly, we find 
$\sigma_{\textrm{H}} = 1/3$ in each of the nine states and inserting
fluxes reveals that three groups of three states each show spectral
flow, with $6\pi=3\times 2\pi$ bringing the system back to the original
point. Parameter sets treated include $V_1/t=100, V_2/t=V_3/t=1$,
$V_1/t=100, V_2/t=V_3/t=2$, $V_1/t=10, V_2/t=V_3/t=2$; 
$t'/t=0.2$ in all cases.

The above results also further ascertain the robustness of the CDW part of
the TPL states as the system size is increased. Taking this robustness
in the strong-$V_1$ limit into account, we can restrict the Hilbert space
to states that contain perfect charge order, as any faults in the charge
pattern will be severely penalized energetically. As detailed in the main
text, the charge order is associated with a 3-fold ground-state degeneracy,
which arises from the translations of the charge pattern. Moreover,
when the strong-$V_1$ limit suppresses deviations from perfect charge
order, it also removes tunneling between the three configurations of
the CDW and the Hilbert space consequently decomposes into three blocks. Treating
each of these blocks corresponds to keeping only the states coming
from one of the copies of the CDW
pattern. The ``pins'' of the pinball can thus be kept fixed and the interacting
``residual'' particles can be treated on the effective honeycomb-lattice model
of Fig.~1(b) of the main text. 

\begin{figure}[t]
\centering
\includegraphics[width=0.7\columnwidth]{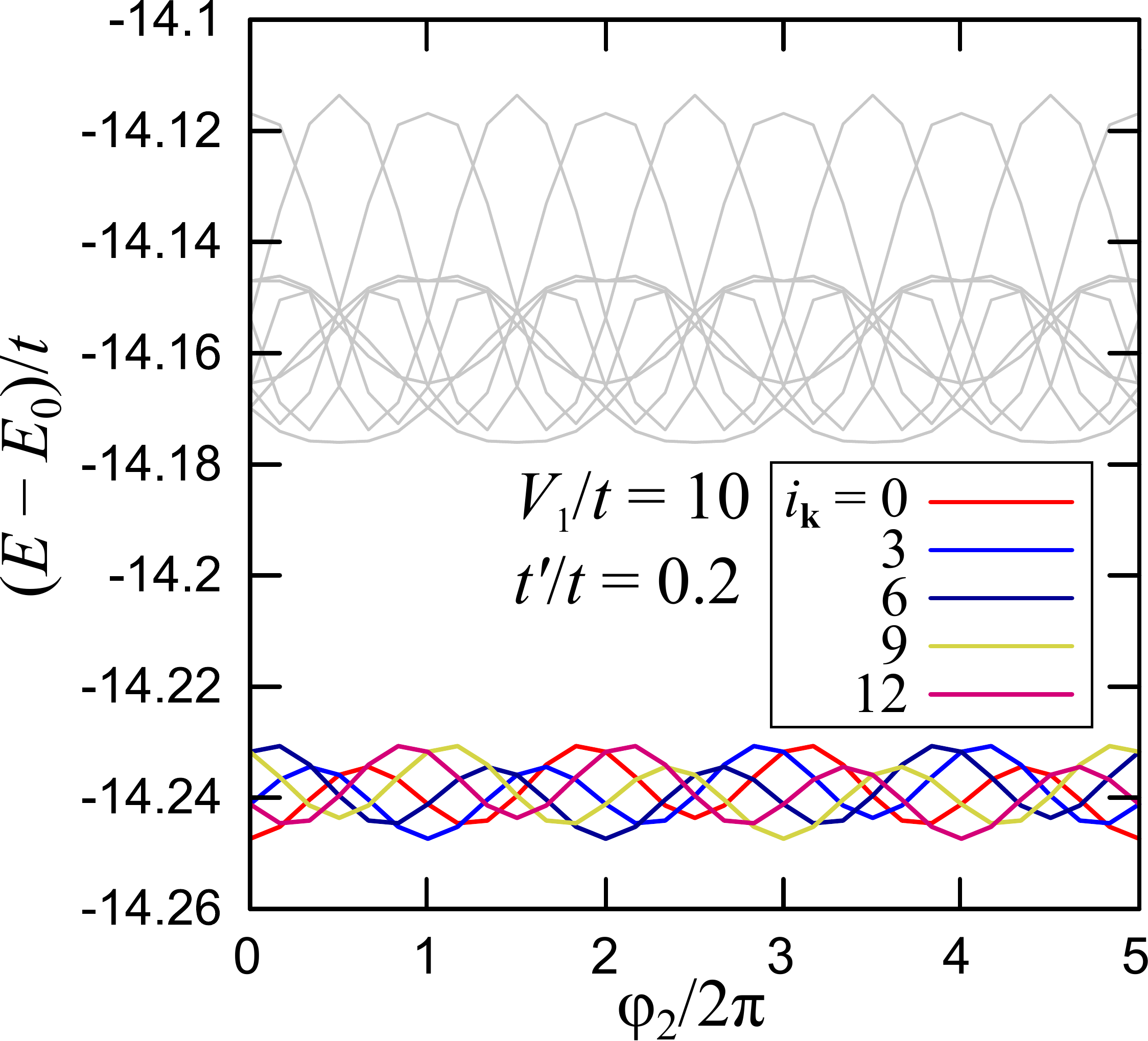}
\caption{Spectral flow of the eigenvalues of the effective honeycomb lattice
[denoted by thick black lines in Fig.~1(b) of the main text] with 6 particles in a $3\times5$-cell
cluster ($\nu=2/5$) with $V_1/t=10$ and $t'/t=0.2$, which models the residual lattice after
assuming perfect charge order.}
\label{fig:hexflux}
\end{figure}

For this model, we have studied the $\nu=2/5$
state for up to 6 particles. We have verified that the ground state is
5-fold degenerate (since we have removed by hand the 3-fold degeneracy of
the CDW), the levels that comprise it exhibit spectral flow (see
Fig.~\ref{fig:hexflux}), and the corresponding states have a very precicely
quantized $\sigma_{\textrm{H}}=2/5$ in units of $e^2/h$. Therefore,
as long as the CDW component of a TPL state remains
robust when the system size is increased (as was seen above), it is reasonable to
expect that TPL states survive for larger systems as well. Phase separation into a charge ordered and an FCI domain within the lattice would be hard to reconcile with the observables discussed above. Additionally, the total energy depending on filling is convex for fillings between 10 and 14 electrons on the 30-site cluster, as shown in Fig.~\ref{fig:phases_V23}(b), also arguing for a thermodynamically stable phase. This
suggests that TPL states will be, in principle, stable in the thermodynamic
limit, at least in a perfect system at zero temperature, since there is no
intrinsic feature of these states that prevents thermodynamic stability.

\end{document}